\documentclass[conference]{IEEEtran}
\IEEEoverridecommandlockouts
\usepackage{cite}
\usepackage{amsmath,amssymb,amsfonts}
\usepackage{algorithmic}
\usepackage{graphicx}
\usepackage{textcomp}
\usepackage{xcolor}
\def\BibTeX{{\rm B\kern-.05em{\sc i\kern-.025em b}\kern-.08em
    T\kern-.1667em\lower.7ex\hbox{E}\kern-.125emX}}
\begin{document}

\title{Inertia constants for individual power plants\\
}

\author{
\IEEEauthorblockN{David Kraljic}
\IEEEauthorblockA{\textit{COMCOM d.o.o.}, \\
\textit{Faculty of Electrical Engineering}, \\\textit{University of Ljubljana}\\
Ljubljana, Slovenia \\
david.kraljic@comcom.si}
\and
\IEEEauthorblockN{Blaz Sobocan \\
Jernej Katanec \\
Matej Logar \\
Miha Troha
}
\IEEEauthorblockA{\textit{COMCOM d.o.o.}\\
Ljubljana, Slovenia}

}

\IEEEoverridecommandlockouts

\maketitle

\IEEEpubidadjcol

\begin{abstract}
Keeping the power system stable is becoming more challenging with the growing share of
renewable energy sources of low or negligible inertia. Inertia constants for individual power plants are generally not known and are roughly estimated by considering the type of power generation technology and the nameplate capacity of plants. More accurate knowledge of inertia constants of individual power plants would give greater transparency to decisions of the
transmission system operator (TSO) and to auctions for the procurement of inertia services. Additionally, a more accurate forecast or estimation of system inertia would
improve the price signals to the power market well in advance of balancing actions taken by the TSO.

We develop methods based on a combination of mathematical optimisation and machine learning that reverse-engineer the inertia constants of individual power plants from the historical values of their power production and from aggregate values of inertia in a power system. We demonstrate the methods for the power system of Great Britain (GB), where historical values for aggregate inertia are published by the TSO. We show that the recovered inertia data is crucial in understanding certain individual balancing decisions by the TSO. We use the reverse-engineered inertia constants to predict system inertia which gives valuable information to the power market.
\end{abstract}

\begin{IEEEkeywords}
Optimisation, Power system dynamics, Power system economics, Power system stability, Reverse engineering
\end{IEEEkeywords}

\section{Introduction}
A satisfactory level of inertia in a power system is crucial for its stability. Inertia is provided by large rotating masses of conventional generators that are electro-mechanically coupled to the grid and can therefore transfer rotational kinetic energy to and from the grid. This has a stabilising effect on the rate of change of frequency (ROCOF) (see e.g. the swing equation \cite{stevenson1994power}), which is crucial in the events of faults, unexpected disconnections and large changes in generation or consumption. Most devices connected to the grid have automatic protection systems that disconnect the device from the grid in case of large ROCOF and thus low inertia is associated with an increased risk of cascading failures \cite{TROVATO2019183}. 

Increasing penetration of renewable generation such as wind and photovoltaics, which are not electro-mechanically coupled to the grid, is causing low levels of inertia and increased instability of the grid \cite{low_inertia_stability}. Therefore, besides increasing the effective grid inertia \cite{virt_inertia}, the estimation of grid inertia is crucial in future-proofing the grid. 

The standard approach to measure inertia, as it was at some point in time, is via frequency transients analysis (e.g. post fault), where the speed and nadir of the frequency drop, or other properties of the transient, are used to extract the amount of grid inertia at the point of the fault\cite{nadir, jap, nord}. The shortcoming of this approach is that the estimation is done `offline' and only estimates the value of inertia at that moment of time in the past. Since faults do not occur regularly this method yields few samples for total system inertia. Another approach is to focus on measuring or estimating the total inertia in real-time (e.g. based on phasor measurement units and/or wide area monitoring \cite{wam, phas, wide}). This approach is valuable to TSOs as it informs operational decisions in real-time, however, such estimates are less useful to the electricity market, where activities usually cease well before real-time.

Instead, we focus in this contribution on the expected or forecast inertia a few hours \emph{before} the delivery start rather than in real-time or offline. The expected or predicted value of inertia for some settlement period a few hours before it starts is important as it determines actions of the TSO that are taken before real-time delivery in anticipation of low inertia. Early inertia prediction can also provide signals to the electricity market, so that mitigating actions can be already taken on the market (and not just in real-time, when it can already be too late, given that large synchronous generators take long to start-up).

The need for procurement of inertia through market mechanisms\cite{pathfinder} e.g. via periodic auctions has appeared because low inertia states due to renewables are becoming more and more frequent. In contrast to approaches in literature that focus on the total system inertia, we estimate inertia values for each individual power plant.
For the transparency and better competition in such markets, it is crucial that individual power plant parameters are known. For example, currently, the Grid Code of GB prevents the TSO to disclose the individual power plant inertia parameters\footnote{https://data.nationalgrideso.com/plans-reports-analysis/covid-19-preparedness-materials/r/transparency\_forum\_inertia\_deep\_dive\_follow-up\_q\&a}.

\begin{figure}[htbp]
\centerline{\includegraphics[width=\columnwidth]{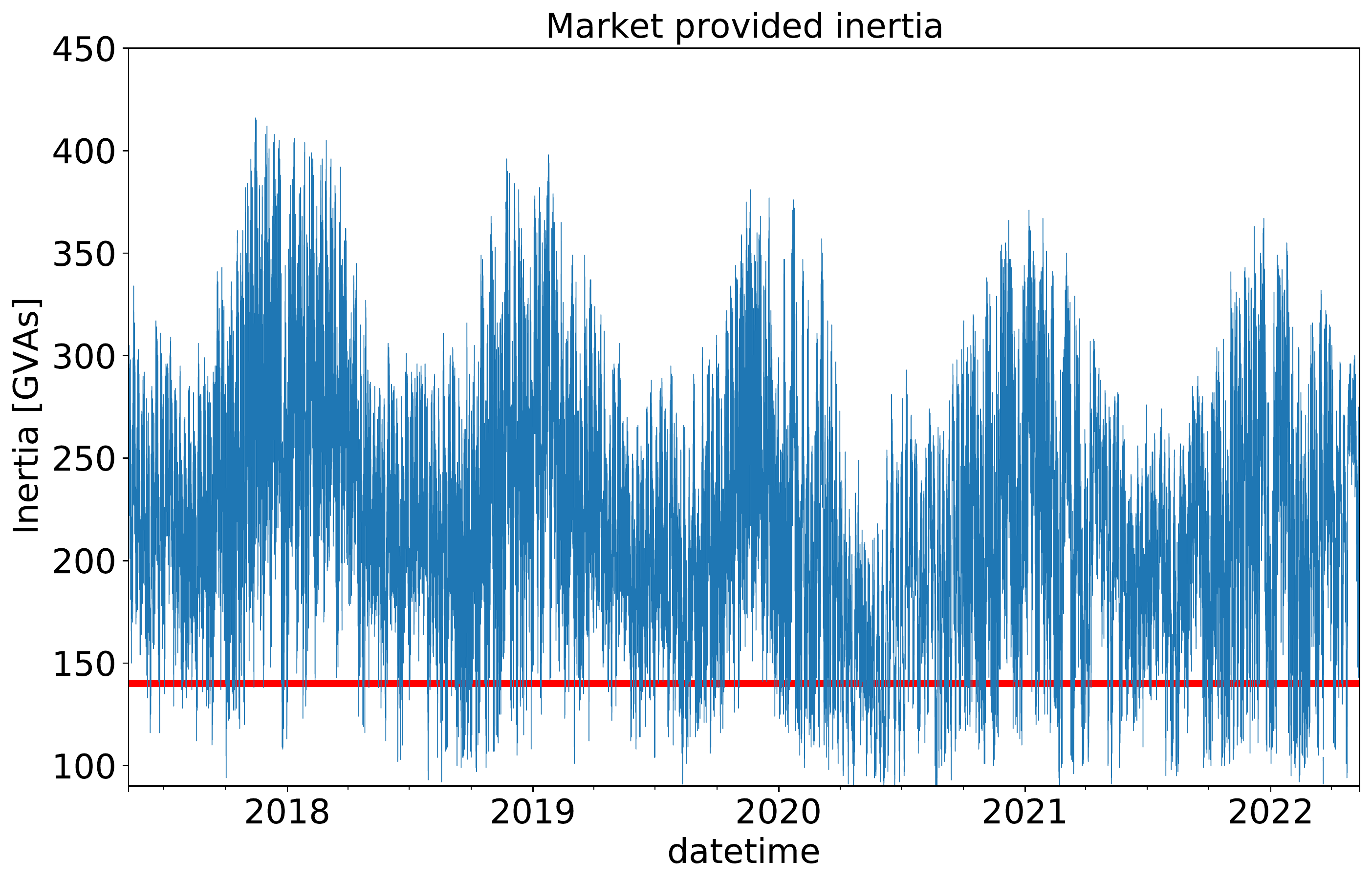}}
\caption{Estimation by the GB TSO for the aggregate `market provided' inertia, which is inertia in the absence of TSO activity. The red line is at 140\,GVAs which is the trigger level for the TSO to take inertia increasing actions.}
\label{fig:market_inertia_history}
\end{figure}

\begin{figure}[htbp]
\centerline{\includegraphics[width=\columnwidth]{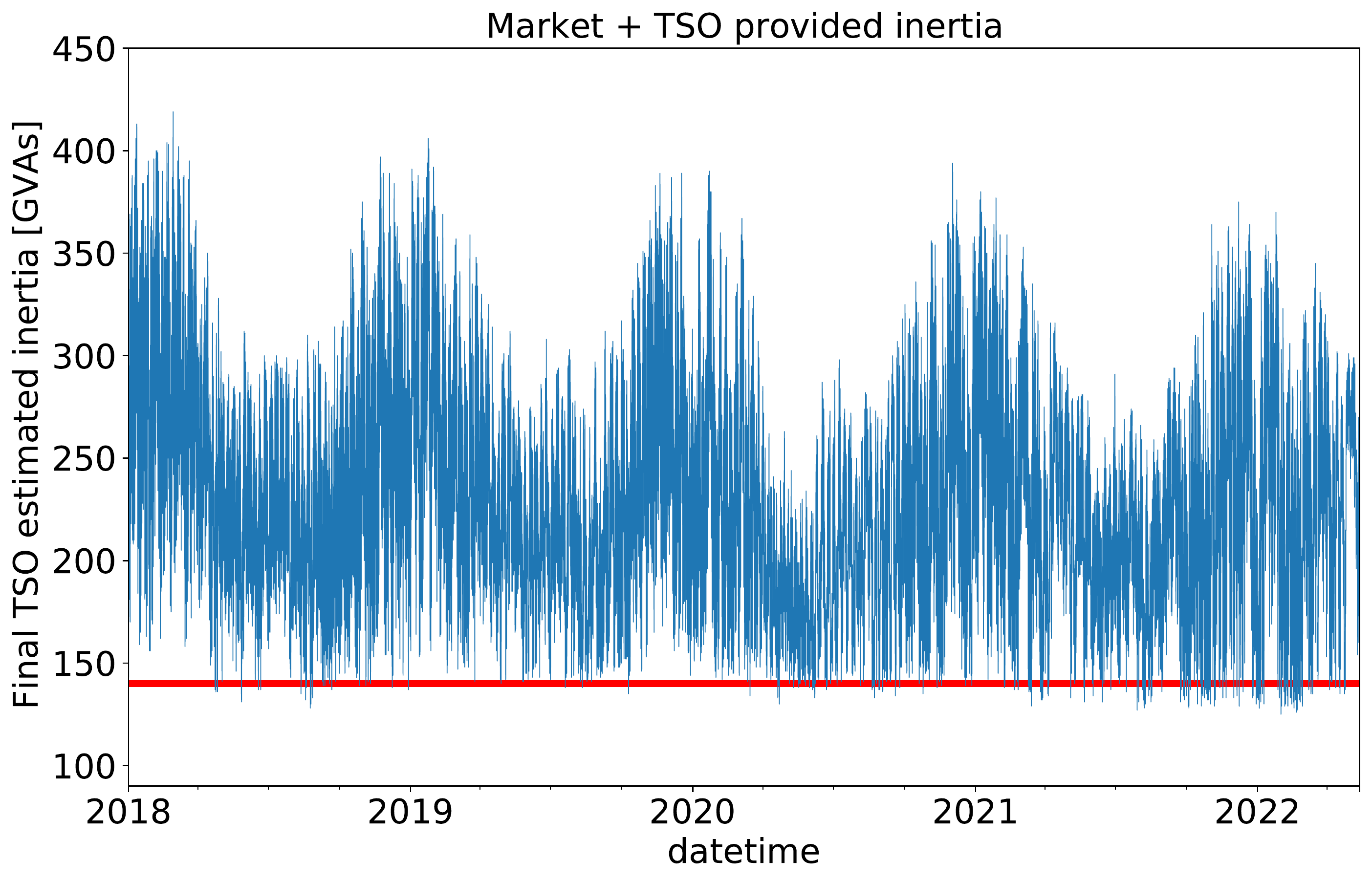}}
\caption{Estimation by the GB TSO for the aggregate `outturn' inertia, which is inertia provided by the market plus actions taken by the TSO. The red line is at 140\,GVAs which is the trigger level for the TSO to take inertia increasing actions.}
\label{fig:market_tso_inertia_history}
\end{figure}

Another area where the data on inertia values for individual power plants would increase transparency is in understanding the individual actions taken by the TSO. Specifically, the GB TSO has a policy to keep grid inertia above 140\,GVAs\cite{140}, as we can see from Figs.\ref{fig:market_inertia_history}, \ref{fig:market_tso_inertia_history}. Therefore, if the predicted aggregate inertia is below this trigger level and the market has the knowledge of individual power plant inertias, it can anticipate which power plants will be run, which is crucial in predicting the imbalance price.

In Section \ref{sec:data} we describe the data used in this contribution, in Section \ref{sec:rec} we describe the method for reconstruction of individual power plant inertias, in Section \ref{sec:res} we present the results, in Section \ref{sec:use} we present use cases of the results and in Section \ref{sec:con} we summarise our contributions and conclude.

\section{Data}
\label{sec:data}
The TSO provides the following datasets:
\begin{itemize}
    \item aggregate amount of inertia provided by the `market' (inertia if TSO did not take any action)\footnote{\label{note1}https://data.nationalgrideso.com/system/system-inertia}
    \item `outturn' inertia, which is the aggregate inertia including the TSO actions\footnotemark[2]
    \item forecasted and historical demand\footnote{https://data.nationalgrideso.com/data-groups/demand}
    \item physical positions of power plants for each settlement period (includes practically every larger transmission connected or embedded power plant)\footnote{\label{note2}https://www.bmreports.com}
    \item TSO balancing actions\footnotemark[4]
\end{itemize}
The `market' and `outturn' inertia values are provided for each settlement period (half-hourly) a few days \emph{after} the settlement. The TSO provides the data on the physical positions of power plants about 1h \emph{before} the settlement period start, whereas the balancing actions are reported in real-time. 

\section{Reconstruction of inertia constants}
\label{sec:rec}
Inertia constants of individual power plants can be reconstructed from the publicly provided data as follows.
We know the aggregate inertia and we know at what level of generation power plants were running in each settlement period. Therefore, the sum of unknown inertias of each running plant in each settlement period has to sum into the known aggregate inertia:
\begin{align}
A^{market}_{t} =& w_{dem.} \cdot d_{t} + \sum_{j} w_{j} \cdot I^{market}_{j, t} \nonumber \\ 
A^{market}_{t+1} =&  w_{dem.} \cdot d_{t+1} + \sum_{j} w_{j} \cdot I^{market}_{j, t+1} \nonumber \\
&\vdots
\label{eq:overdetermined}
\end{align}
where $A^{market}_t$ is the aggregate inertia provided by the market at time $t$ and $I^{market}_{j, t}$ is an $1/0$ indicator whether power plant $j$ is turned on at time $t$, $w_{j}$ is the inertia constant of power plant $j$, $d_t$ is the demand in the system (in GW) and $w_{dem.}$ is the inertia of demand in the units of GVAs.

The set of linear equations \eqref{eq:overdetermined} is overdetermined and cannot be solved exactly in practice, because we might not take into account everything that is used to estimate aggregate inertia by the TSO. A classic resolution to the overdetermined problem is to find an approximate, best-fit solution, which is usually done by Ordinary Least Squares. However, because the inertia constants $w_{j}$ are non-negative we need to solve a constrained problem which leads us to the following quadratic programming problem:
\begin{align}
\underset{w}{\text{argmin}} \sum_t \left( A^{market}_t - \left(w_{dem.} \cdot d_{t} + \sum_{j} w_{j} \cdot I^{market}_{j, t}\right)\right)^2 \nonumber \\
\quad \text{such that }  w_{j} \geq 0  \land w_{dem.} \geq 0 \quad \quad \quad
\label{eq:quadratic}
\end{align}

The data on power plants contains all types of plants and participants in the Balancing Mechanism; for some of them we know (or strongly suspect) that inertia constants are zero (wind, solar, battery). We would like to encourage the quadratic program \eqref{eq:quadratic} to set those values to zero but without forcing it. For this purpose, we use regularisation, a method in machine learning, by adding a penalty term to the objective in \eqref{eq:quadratic} to make the problem of this type better behaved and encourage the coefficients $w_j$ to be zero if they do not contribute to the quality of the fit. Specifically, we add a  $l_0$ norm term $ \lambda \sum_j \| w_j \|_0$ to the objective in \eqref{eq:quadratic} which penalises the number of nonzero values of $w_j$ with strength $\lambda$, but does not `shrink' (or bias) coefficients towards zero. This way the nonzero reconstructed inertia constants are not smaller that they should be.

\subsection{Including the TSO actions}
The data available from the GB TSO also includes the aggregate inertia caused by the actions of the TSO. We can merge this with the data for the `market provided' inertia to double the dataset (for each settlement period we have two independent datapoints then) in order to improve the best-fit solutions for $w_j$. However, the TSO instructs only a few power plants in each settlement period and therefore we expect that the improvement in reconstructed values from adding the second dataset (TSO actions) will not be dramatic.

The full problem we solve is as follows:
\begin{align}
\underset{w}{\text{argmin}}\sum_t \Bigg[ \left( A^{market}_t - \left(w_{dem.} \cdot d_{t} + \sum_{j} w_{j} \cdot I^{market}_{j, t}\right)\right)^2 \nonumber \\
+ \left( A^{TSO}_t - \sum_{j} w_{j} \cdot I^{TSO}_{j, t}\right)^2 \nonumber \\
+ \lambda \sum_j \| w_j \|_0
\Bigg] \nonumber \\
\quad \text{such that }  w_{j} \geq 0  \land w_{dem.} \geq 0 \quad \quad \quad
\label{eq:full_problem}
\end{align}
where  $A^{TSO}_t$ is the aggregate inertia as a consequence of the TSO actions and $I^{TSO}_{j, t}$ is a 1/0/-1 indicator whether the TSO action switched the power plant $j$ ON (1), OFF (-1), or did not instruct the power plant (0) at time $t$.

\section{Results}
\label{sec:res}
We solve the optimisation problem \eqref{eq:full_problem} using the optimisation software \textsc{gurobi} \cite{gurobi}. The data can also be organised such that the problem \eqref{eq:full_problem} can be solved with the open-source python package \textsc{scikit-learn}\cite{scikit-learn}. Note, however, that the package does not implement the $l_0$ penalty, so either we do not regularise (unstable solution) or we use the implemented $l_1$ penalty (slightly reduces reconstructed inertia constants).
\subsection{Predictive performance}
To evaluate the predictive performance of our approach we split the data into train and test sets, with train set encompassing years 2020 to 2021 and the test set the first four months of 2022. On the train set we extract the individual power plant inertia values by solving \eqref{eq:full_problem}. We predict aggregate inertia provided by the market (as is estimated by the TSO) from the data of expected physical positions of power plants: we multiply the indicator whether the power plant will run or not with the reconstructed inertia value from the train set and finally sum all contributions. We compare this value with the TSO published estimate.

\begin{figure}[htbp]
\centerline{\includegraphics[width=\columnwidth]{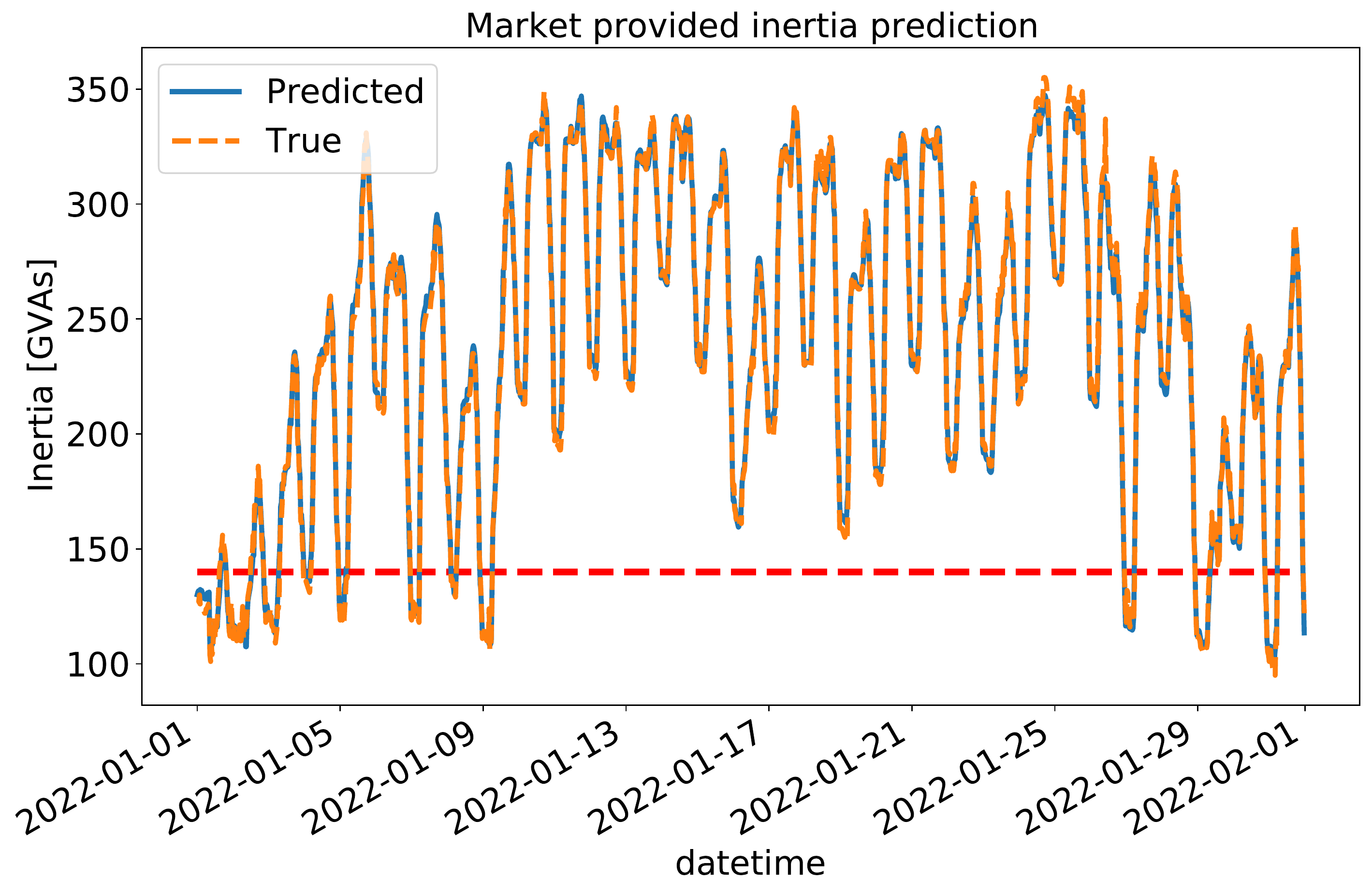}}
\caption{Aggregated inertia provided by the market for January 2022 as predicted from reconstructed inertia values and physical positions of individual power plants and as reported by the TSO (`True'). We also mark the 140GVAs line that triggers the TSO to take corrective actions to increase inertia.}
\label{fig:market_inertia_prediction}
\end{figure}

We reach the Mean Absolute Error for the aggregate inertia on the test set of 3.5\,GVAs, which is roughly the inertia of a single power plant. Comparing this value to the typical value for aggregate inertia of 200\,GVAs we reach Mean Absolute Percentage Error of 1.7\%. The results are illustrated in Fig.\ref{fig:market_inertia_prediction}.

\subsection{Consistency of reconstructed values}
Inertia values for the power plants in the GB grid are not known publicly. However, we found data \cite{ng_inertia_data} that overlapped with our dataset for the case of two pumped storage units\footnote{Dinorwig Power Station, Dinorwig, Wales, UK}. The values we obtain are 1.48\,GVAs and 1.54\,GVAs whereas the values as reported in \cite{ng_inertia_data} are 1.59GVAs and 1.59GVAs respectively.

Another approach to validate our results, in the absence of true data for individual plants, is to compare whether our reconstructed values fall in the ranges as reported elsewhere in the literature. The values are usually reported in terms of the inertia constant H measured in seconds, which is the raw inertia value (as reconstructed in this article) normalised by the nameplate capacity of power plants.

The values we obtain for thermal generation (CCGT, coal, biomass, gas) are in the range (3-10\,s) as suggested by literature \cite{inertia_values_1, inertia_values_2, inertia_values_3, inertia_values_4}, however the values for hydro power plants are a bit higher. We attribute this to the fact that the power plants we consider have smaller ratings (mean is 45MW) and might have higher inertia than literature values (2-4\,s) because the inertia constant typically scales inversely with the power rating\cite{inv_inert}.

{The limitation of the methods proposed in Sec.\,\ref{sec:rec} is that certain power plants that are very similar in physical characteristics (e.g. units at the same site) have very similar physical positions and are almost always instructed together by the TSO. These power plants make the optimisation problem unstable (colinearity) and the values of their reconstructed inertias is not well constrained. These problems can be ameliorated at the data processing stage, where power stations with near identical physical positions are assumed to have the same inertia, which is encoded as an additional constraint in the optimisation problem \eqref{eq:full_problem}.}

\begin{figure}[htbp]
\centerline{\includegraphics[width=\columnwidth]{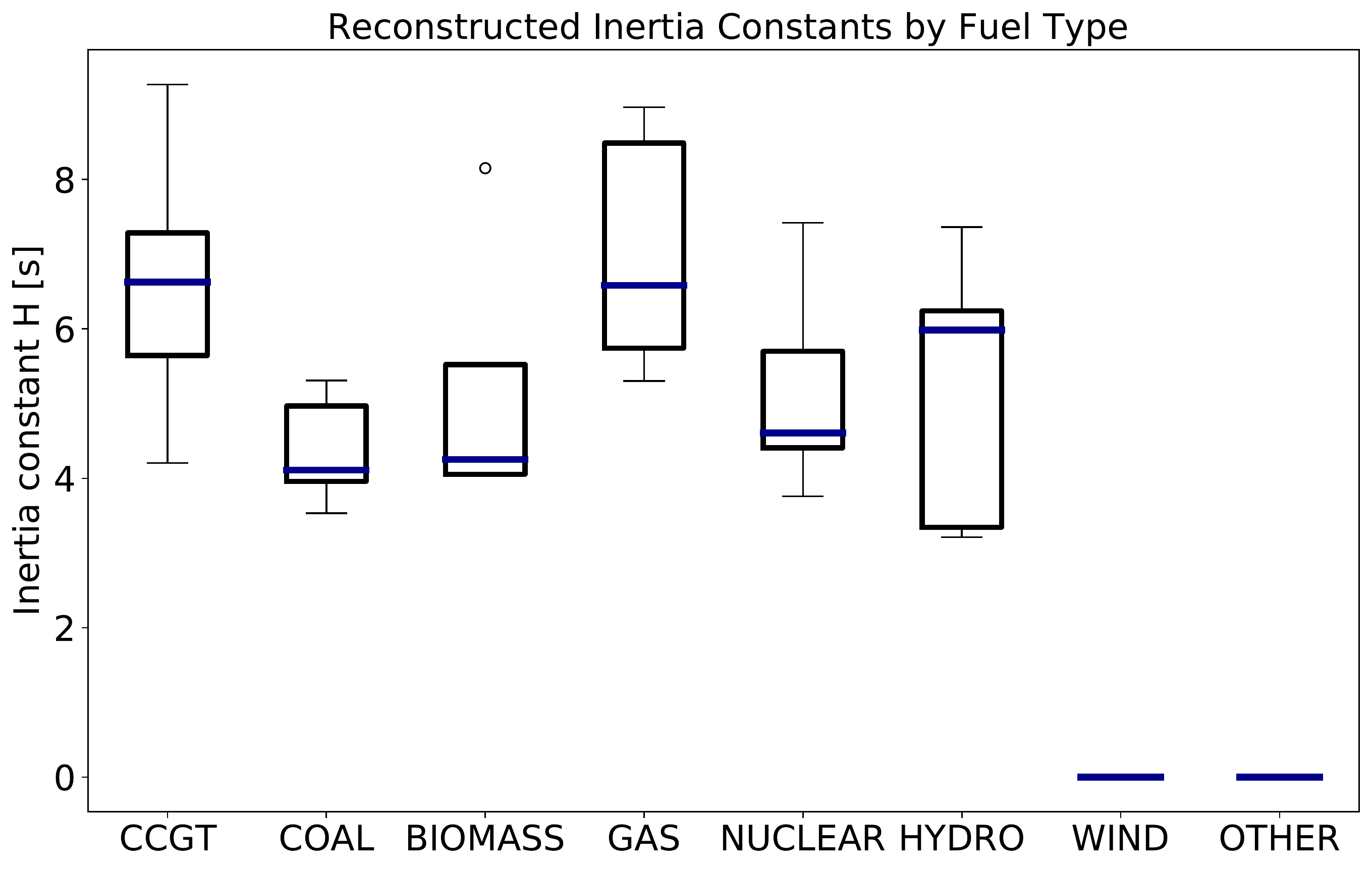}}
\caption{Reconstructed inertia constants H in units of seconds sorted by fuel type. Note that solving \eqref{eq:full_problem} reconstructs the zero inertia for all wind power plants as well as for `other' (battery, virtual, ...) types of plants. Also note that the reconstructed inertias by fuel type are in the ranges as expected from the literature.}
\label{fig:reconstructed_inertia_fuel}
\end{figure}

\section{Use cases}
\label{sec:use}
\subsection{TSO action anticipation}
The GB TSO provides the data on each power plant dynamic parameters (ramping rates, stable production levels, maximum production levels, minimum notice to start, minimum zero times, minimum nonzero times, ...) as well as the prices for instructing each of the power plants\footnote{https://www.bmreports.com/bmrs/?q=balancing/}. Thus, if the situation of low inertia arises (as signalled by our predictions) that would trigger the TSO into deliberate actions with the sole purpose of increasing inertia, we can predict or anticipate these specific actions.

Assuming the TSO behaves in a cost effective way, we need to solve the following optimisation problem: minimise the cost of starting up (or keep running) of power plants such that the total inertia in the grid rises above the minimum level (140GVAs), while ensuring that dynamic parameters are respected (notice time to start, ramp rate, minimum zero time) and also ensuring the energy balance of the grid as well as other operational needs (constraints, voltage, reserve).
This is a complicated problem as it includes precise modelling of power plant dynamics and the power system. Focusing on just the inertia subproblem we can sketch it:

\begin{align}
\underset{I^{TSO}}{\text{argmin}} \sum_j C_{j, t} \cdot I^{TSO}_{j, t} \nonumber \\
\text{such that} \nonumber \\
\sum_j w_j \cdot I^{TSO}_{j, t} \geq 140 \text{\,GVAs}\nonumber \\
\text{minimum notice to start is respected } \forall j \nonumber \\
\text{ramping rate is respected } \forall j\nonumber \\
\label{eq:actions_predictions}
\end{align}
where $C_{j, t}$ is the minimum cost to start up or keep running power plant $j$ at time $t$, and as before $I^{TSO}_{j, t}$ signals if the TSO action is instructing a power plant to run or not and $w_j$ is the inertia of power plant $j$.

This subproblem is expected to yield good results when other TSO considerations are not important (energy, reserve, constraints, voltage). We selected a period of exceptionally low `market' inertia (predicted 109.9\,GVAs, TSO estimated 108\,GVAs)  and the excess of zero inertia renewable energy as well as no active constraints to manage (settlement day 09-01-2022, settlement period 8, lasting from 3.30am to 4.00am). This ensured that the only operational concern for the TSO was the inertia subproblem \eqref{eq:actions_predictions}. 

We solved the optimisation problem \eqref{eq:actions_predictions} for the expected TSO actions and obtained 7 power plants that should keep running and 1 power plant that should switch on. Power plants that should keep running have ids\footnote{https://www.bmreports.com/bmrs/?q=balancing/searchbyBMUnit}: T\_CARR-1, T\_CARR-2, T\_PEMB-31, T\_STAY-1, T\_WBURB-1, T\_WBURB-2, T\_WBURB-3. These power plants have each 3.4\,GVAs of reconstructed inertia and were indeed selected by the TSO to keep running, despite an overall excess of wind power\footnote{https://www.bmreports.com/bmrs/?q=/balancing/detailprices/2022-01-09/8/sysprices}. Our solution also yielded power plant T\_COSO-1, with reconstructed inertia of 7.7\,GVAs, that should turn on.
However, the GB TSO did not take such action and start this power plant. The actual aggregate inertia after the TSO actions was 132\,GVAs, which is below its own target of 140\,GVAs, and this is the reason our optimisation problem selected one additional power plant (to reach 140\,GVAs as specified in \eqref{eq:actions_predictions}).

\subsection{Market price formation}
Usually, the participants in the electricity market are only concerned with the energy balance, that is, matching the supply and demand, whereas the `system' issues, such as transmission constraints, reserve requirements, and inertia are left to the TSO to manage, usually in a separate balancing market.
The ability to precisely predict the inertia provided by market (i.e. before TSO actions) is valuable as it sends signals to the market of the operational situation and potential TSO actions before they happen. This might help to form fairer prices. Typically, in a low inertia situation, usually because of the high renewable production, the TSO instructs conventional synchronous generators to switch on, which they do at prices higher than the market (since they were switched off as market prices were too low). These costs incurred by the TSO then have to be collected through imbalance prices and other charges. 
Imbalance prices are a key driver of the market price\cite{imbalance_chasing}. Therefore, if the market gets the information that inertia is too low, which will trigger the TSO into expensive actions, the market will move towards higher prices anticipating higher imbalance price, which in turn motivates extra controllable generation to run and thus ameliorates the operational situation of low inertia. 

\section{Conclusions}
In this contribution we presented a method for reverse engineering inertia values of individual power plants from data on the aggregate inertia and on the power plant physical positions. We demonstrated the method on the GB grid for which the necessary data is available. We show that the values extracted are consistent with literature estimations based on the power plant type and with a few individually known values. We also show that the extracted values can be used to predict the aggregate inertia values as estimated by the TSO to great accuracy. We show a method to predict individual TSO actions related to inertia and discuss in what situations the method is suitable. These methods and predictions significantly increase transparency of the electricity market and will become more and more relevant in the future as electric grids move towards frequent situations of low inertia. 

Future work will include predicting the market provided inertia much earlier (in this contribution we predict 1h before delivery) from various data sources (not just physical positions of power plants). We will integrate the inertia subproblem into the wider TSO operational optimisation problem, such that the TSO actions regarding inertia can be predicted even when there are other operational concerns (e.g. constraints, reserve) and not just in simple cases when the power system need is purely more inertia.

\label{sec:con}

\section*{Author contributions}
Conceptualisation, model training and writing (DK). Use cases and data infrastructure (BS). Literature overview (ML). Data scraping and processing (JK). Writing (MT).

\end{document}